\documentclass{article}
\newcommand{\leftragged}{\raggedright}
\usepackage[preprint]{neurips_2020}
\usepackage{geometry}
\usepackage{graphicx}
\usepackage{qtree}
\usepackage{tabularx}
\usepackage{ragged2e}
\usepackage{booktabs}
\usepackage{siunitx}
\usepackage{ragged2e}
\usepackage{rotating}
\usepackage{adjustbox}
\usepackage{caption}
\usepackage[T1]{fontenc}
\usepackage{pdflscape}
\usepackage{nameref}
\usepackage{pdfpages}
\usepackage[multiple]{footmisc}
\usepackage{float}
\newcolumntype{R}[1]{>{\RaggedRight}p{#1}}
\usepackage{amssymb}
\usepackage[x11names,dvipsnames]{xcolor}
\usepackage{verbatim}
\usepackage{subcaption}
\usepackage{array}
\usepackage{natbib}
\usepackage{hyperref}
\usepackage{mathtools}
\usepackage{xspace}
\usepackage{cleveref}
\usepackage{svg}
\usepackage{transparent}
\usepackage[toc,title,page]{appendix}
\definecolor{orangefull}{RGB}{230, 159, 0}
\newcommand*{\eg}{e.g.\@\xspace}
\newcommand*{\ie}{i.e.\@\xspace}

\usepackage{siunitx}
\usepackage[pagewise]{lineno}  
\colorlet{orange}{orangefull!20!white}
\definecolor{bluefull}{RGB}{86, 180, 233}
\colorlet{blue}{bluefull!20!white}
\definecolor{green}{RGB}{0, 158, 115}
\colorlet{green}{green!20!white}

\title{Uncertainty in wind and solar projections depends on global and regional climate models}

\author{%
  Nina Effenberger \\
  Institute for Atmospheric and Climate Science\\
  ETH Zürich\\
  \texttt{nina.effenberger@env.ethz.ch} \\
  \And
    Reto Knutti \\
  Institute for Atmospheric and Climate Science\\
  ETH Zürich\\
}

\begin{document}

\maketitle

\begin{abstract}
Ensembles of regional–global climate model combinations show substantial spread in projected wind and solar resources. Using 31 RCM–GCM pairs, we quantify the sources of this spread with a spatially and seasonally resolved variance decomposition, separating contributions from RCMs and GCMs. For both wind speed and solar radiation, RCMs dominate the variability in the absolute historical fields. In contrast, projected changes in wind speed are largely controlled by the driving GCMs, except in mountainous regions where RCM-induced variance becomes larger than that induced by GCMs. For solar radiation, contributions are strongly season-dependent, with RCMs dominating in summer and GCMs in winter. Our findings support that GCM and RCM variability together define the uncertainty of wind and solar climate projections. This provides guidance for designing climate model ensembles that better support uncertainty-aware energy system decisions under climate change.
\end{abstract}

\section{Introduction}
The plan to expand renewable energy in climate mitigation strategies increases the need for robust information on how resources may evolve under climate change. However, projected changes in wind and solar resources remain highly uncertain. Previous studies show that signals of projected changes in wind power production are often weak, strongly seasonal, and sensitive to model choice. For example, \citet{reyers2016future} report substantial variation in wind energy projections across CMIP5 models, including opposite signs of trends, highlighting the need for multi-model ensemble analyses to better quantify and understand future changes. Similarly, \citet{carvalho2021wind} identify substantial uncertainty in seasonal wind power changes across different emission pathways. For solar power, \citet{jerez2015impact} demonstrate that uncertainty in photovoltaic output projections is also high, again with limited agreement on both the magnitude and direction of change.

Uncertainty in climate projections has long been known to arise from multiple sources. \citet{hawkins2009potential} formalize a decomposition of projection uncertainty into internal variability, model uncertainty, and scenario uncertainty, and show that the balance of these components depends on variable, region, and forecasting horizon. \citet{knutti2013robustness} further conclude that inter-model spread remains large even in newer generations of climate models, underscoring that improved physics alone does not necessarily reduce projection uncertainty. While emergent constraints have successfully reduced some uncertainties \citep{hall2019progressing}, performance-based model weighting remains challenging \citep{merrifield2023climate, knutti2010challenges}, and uncertainties therefore remain large in many cases. Applied to renewable resources, this implies that uncertainties in wind and solar projections are not only substantial but also stem from distinct and interacting components within the modeling chain. \citet{reyers2016future} already point to the challenge of identifying the drivers of these uncertainties in energy-related variables in CMIP5 models. 

To obtain more detailed local information, researchers increasingly rely on high-resolution data. In this context, high-resolution regional climate models (RCMs) are often assumed to add value to wind and solar projections, especially in complex terrain. Yet, evidence is mixed. While \citet{herrmann2011representation} and \citet{rummukainen2016added} demonstrate enhanced representation of near-surface winds at higher resolution in mountainous areas, \citet{di2015challenges} and \citet{morelli2024climate} find that added value is not always guaranteed when spatial resolution is increased. Comprehensive evaluations, such as those by \citet{vautard2021evaluation}, further show that RCMs can improve specific variables and regional features, though their influence on projected changes remains difficult to generalize. As a result, it remains unclear whether and when regional climate model data are required for wind and solar power assessments. Moreover, it is also unclear whether uncertainty in projected changes in wind and solar resources primarily stems from the driving global climate models (GCMs) and how much is introduced by the RCMs. Compounding this uncertainty is the issue of model dependence. Structural similarities within 'families' of models can lead to shared biases, causing the effective ensemble spread to be narrower than it appears \citep{merrifield2023climate, abramowitz2015climate, knutti2013climate}.

To address this gap, we quantify the relative contribution of different sources to climate model uncertainty. We use ANOVA-based methods to separate the roles of GCMs and RCMs within the coordinated modeling framework CORDEX \citep{evin2021balanced, christensen2020partitioning}. Applying such approaches to wind and solar projections helps clarify whether uncertainties are primarily driven by large-scale circulation differences in the GCMs or regional processes captured in the RCMs. This work uses such a decomposition to provide a spatially and seasonally resolved assessment of uncertainty in raw data and projected changes in wind and solar resources. We further underscore that the absence of clear trends does not imply a lack of underlying change. 

\section{Methods}
In this study, we apply a spatially and seasonally resolved variance-decomposition framework to attribute uncertainty in wind speed and solar radiation projections to the driving GCMs and nested RCMs. By analyzing multi-model ensembles, we quantify contributions to both historical fields and projected changes and identify dominant sources of variability across regions and seasons. 

\subsection{Data}
\begin{figure}
    \centering
    \includegraphics[width=\linewidth]{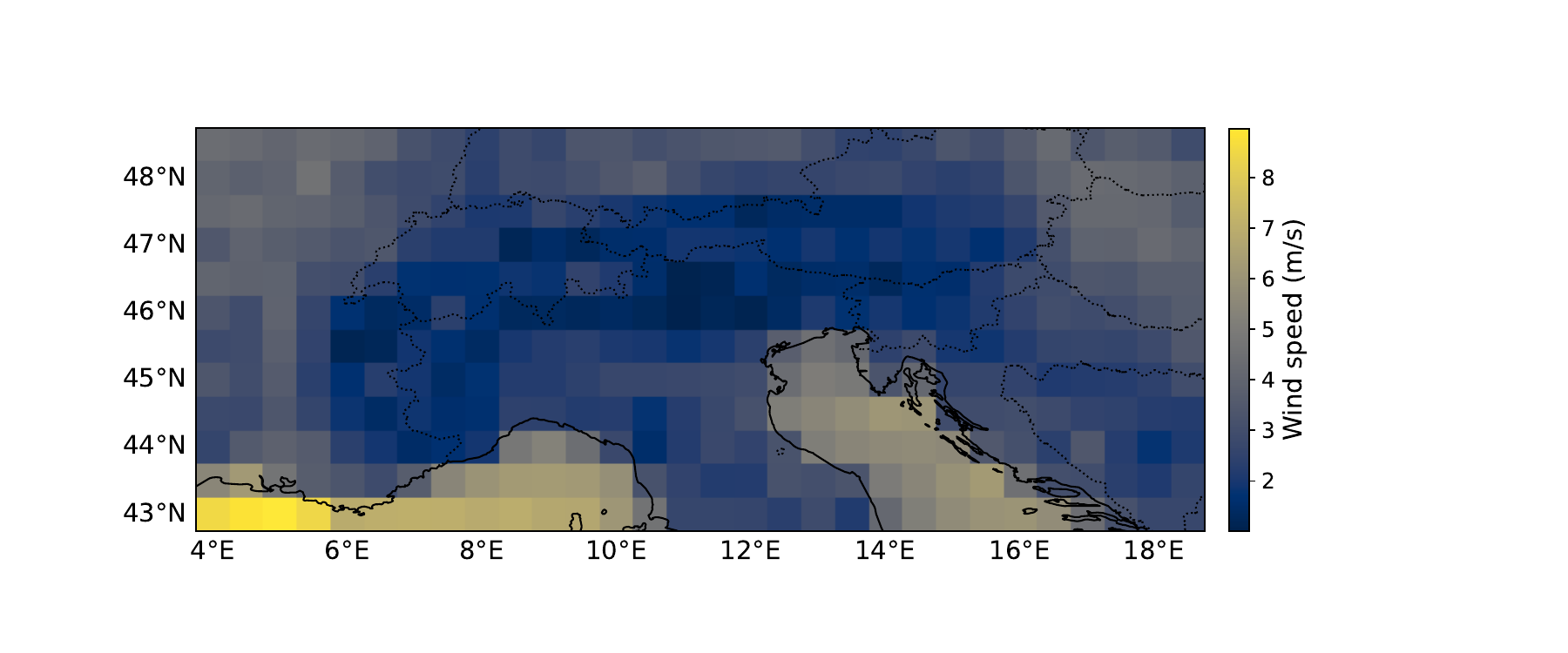}
    \caption{Geographic region analyzed in this study, covering the Alps and surrounding areas in Central Europe.}
    \label{fig:region}
\end{figure}

Regional climate data are from CMIP5-driven EURO-CORDEX simulations of RCP8.5, using a single realization per model; these are provided at a three-hourly average resolution, bilinearly remapped to a uniform $0.5^\circ$ grid, and converted to a 360 day calendar. The analysis uses the 31 available RCM–GCM pairs shown in \Cref{tab:overviewdata}, covering six GCMs and six RCMs over the Alpine region (\Cref{fig:region}) as defined in \eg \citet{pepin2022climate}.  

\begin{table}[h]
    \centering
    \small
    \caption{Overview of the EURO-CORDEX RCM–GCM ensemble. Available simulation pairs are marked with \checkmark, and missing combinations are indicated by  $\times$.}
    \label{tab:overviewdata}
\begin{tabularx}{\textwidth}{l|*{6}{>{\centering\arraybackslash}X}}
\hline
RCM $\backslash$ GCM 
& CNRM-CERFACS-CNRM-CM5 
& ICHEC-EC-EARTH 
& IPSL-IPSL-CM5A-MR 
& MOHC-HadGEM2-ES 
& MPI-M-MPI-ESM-LR 
& NCC-NorESM1-M \\
\hline
CLMcom-ETH & \checkmark & \checkmark &$\times$& \checkmark & \checkmark & \checkmark \\
CNRM       & \checkmark & $\times$&$\times$& \checkmark & \checkmark & \checkmark \\
DMI        & \checkmark & \checkmark & \checkmark & \checkmark & \checkmark & \checkmark \\
KNMI       & \checkmark & \checkmark & \checkmark & \checkmark & \checkmark & \checkmark \\
MOHC       & \checkmark &$\times$&$\times$& \checkmark & \checkmark & \checkmark \\
SMHI       & \checkmark & \checkmark & \checkmark & \checkmark & \checkmark &\checkmark  \\
\hline
\end{tabularx}
\end{table}

\subsection{Definition of Aspects}
We analyze six aspects of the solar radiation and wind speed fields. In this part, we focus on the aspects themselves, and in later sections, we describe the analyses performed. We provide an overview of the aspects in \Cref{tab:overview}. For all analyses, we split the 10 years into seasons, defined as winter (December, January, February), spring (March, April, May), summer (June, July, August), and autumn (September, October, November).  
As a first indicator, we compute the mean for the respective season and time period at each grid point and for each model. 

The data for each model are spatio-temporal, and \(X_{m,n,t}\) denotes the value of a climate variable of a model at grid point lon=$m$, lat=$n$, and time $t$.  
Let \(T_s\) be the set of time indices corresponding to season \(s\) and the chosen time period.  

The seasonal mean at each grid point for each model is

\begin{equation}
\bar{X}_{m,n}^{(s)} = \frac{1}{|T_s|} \sum_{t \in T_s} X_{m,n,t} \text{,}
\end{equation}

where \(|T_s|\) is the number of time steps in season \(s\). Similarly, we compute the temporal variance at all grid points 

\begin{equation}
\sigma_{m,n}^{2\,(s)} = \frac{1}{|T_s|} \sum_{t \in T_s} \Big( X_{m,n,t} - \bar{X}_{m,n}^{(s)} \Big)^2 \text{.}
\end{equation}

To better characterize the distribution, we compute both the 90th percentile and the median at each grid point. Zero values in the solar radiation data are excluded to avoid zero medians in winter, which would prevent performing ANOVA at some locations.

We then extend the analysis to multivariate aspects, where we compute co-occurring lows and spatial correlation. To compute co-occurring lows, additional preprocessing is required. We first average the data to daily values. The 20th percentile is then computed separately for the daily averages of solar radiation data and wind speeds per season and dataset. We then compute the number of days per season during which both values are in the lowest 20\% simultaneously marking days that are potentially demanding for the energy system. A similar methodology was adopted by \citet{kapica2024potential} and \citet{effenberger2025bridging} to investigate compound renewable energy droughts and we hereafter refer to them simply as \textit{droughts}.

As a second multi-variate aspect, we consider gridpoint-wise correlation between the two variables. For that, we first divide the data into seasons again, and then compute seasonal wind and solar radiation anomalies by subtracting the seasonal mean. At each grid point, we then calculate the Pearson correlation of these anomalies along the temporal dimension, producing spatial fields that indicate where increases and decreases in wind tend to coincide with increases and decreases in solar, respectively. Correlation coefficients are Fisher z-transformed to stabilize variance and to make them appropriate for subsequent ANOVA.

\begin{table}[]
    \centering
    \caption{Overview of the aspects analyzed. The first four rows show the single variable aspects and the two last rows the multivariate aspects.}
    \begin{tabularx}{\textwidth}{l|*{2}{>{\leftragged\arraybackslash}X}}
    \hline
         Aspect & Relative contribution of RCM/GCM choice to (change of)...   \\ \hline
         Temporal mean & ... temporal mean at each location \\
         Temporal variance & ... temporal variance at each location \\
         90th percentile & ... high extremes at each location   \\
         50th percentile & ... median values at each location \\
         Droughts & ... number of days with simultaneously low wind speed and solar radiation at each location\\
         Spatial correlation & ... correlation between wind speed time series and solar radiation time series at each location\\
         \hline

    \end{tabularx}
    \label{tab:overview}
\end{table}
\subsection{Analysis Approach}
After computing the six aspects for the two time periods 1995–2004 and 2045–2054, our main analysis proceeds in two steps. First, we compute variability contributions across climate models for each aspect of the raw climate data from 1995–2004, revealing the sources and magnitude of uncertainty in historical fields. Second, we repeat the same analysis for the climate change signal, defined as the difference between the future (2045–2054) and historical (1995–2004) periods for each aspect. To calculate variability contributions, we use ANOVA, which we describe in \Cref{sec:anova}.

When performing the ANOVA on the raw data, we apply it to the values described in \Cref{tab:overview}. To quantify changes between future and historical periods, we use the Mean Signed Percentage Error (MSPE)

\begin{equation}
\text{MSPE} = \frac{100}{n} \sum_{t=1}^{n} \left( \frac{y_t^\text{f} - y_t^\text{h}}{y_t^\text{h}} \right) \texttt{,}
\end{equation}

where $y_t^\text{h}$ and $y_t^\text{f}$ are the historical and future values. MSPE expresses proportional change relative to the historical period, enabling robust comparisons across regions and variables. This choice ensures that all results are interpretable as a mean temporal change per location, normalized to a percentage scale. As MSPE can become unstable when historical values are very small, such as for solar radiation in winter, we use absolute errors for changes in mean solar radiation

\begin{equation}
\text{MAE} = \sum_{t=1}^{n} ({y_t^\text{f} - y_t^\text{h}})\text{.}
\end{equation}

For correlation-based aspects, we apply a Fisher z-transformation to stabilize variance, allowing correlations to be analyzed consistently with MSPE in the ANOVA framework.

\paragraph{Robustness} While the full ensemble analysis is presented in the main text, we also perform a sensitivity check to ensure our findings are robust towards the choice of models used. Specifically, we apply two complementary selection methods using ERA5 as the reference dataset to confirm that the results are not artifacts of the primary data-selection process. First, we select the 15 models with the highest correlation to ERA5 (computed point-wise in space and time). Second, we average over time and select the 15 models with the lowest average MSPE across space relative to ERA5. These two criteria yield different subsets of the 31 available models covering almost all datasets, compare \Cref{tab:dataselection}; yet all main conclusions remain consistent across both selections. We report the corresponding results for the two sub-selected datasets in \Cref{sec:subset}.

\subsection{Analysis of variance}
\label{sec:anova}
To quantify the relative contributions of global climate models (GCMs) and regional climate models (RCMs) to the variability of climate aspects, we apply a two-way analysis of variance (ANOVA), which has been used in similar studies before \citep{yip2011simple}. Let $Y_{ij}$ denote the value of a climate aspect (e.g., mean spatial field or 90th percentile) for the $i$-th GCM and $j$-th RCM. The two-way ANOVA model is expressed as:

\begin{equation}
    Y_{ij} = \mu + \alpha_i + \beta_j + \varepsilon_{ij},
\end{equation}

where $\mu$ is the overall mean, $\alpha_i$ is the effect of the $i$-th GCM, $\beta_j$ is the effect of the $j$-th RCM . The residual term $\varepsilon_{ij}$ therefore includes GCM–RCM interaction effects as well as other unexplained sources of variability. This choice is motivated by the incomplete GCM–RCM matrix (\Cref{tab:overviewdata}) and the focus on first-order contributions to uncertainty. We discuss this in \Cref{sec:discussion}. 

Accordingly, the total variability is decomposed as
\begin{equation}
\label{eq:ss_decomposition}
\mathrm{SS}_{\mathrm{total}} =
\mathrm{SS}_{\mathrm{GCM}} +
\mathrm{SS}_{\mathrm{RCM}} +
\mathrm{SS}_{\mathrm{res}},
\end{equation}
where the residual sum of squares is given by
\begin{equation}
\label{eq:ss_residual}
\mathrm{SS}_{\mathrm{res}} =
\mathrm{SS}_{\mathrm{interaction}} +
\mathrm{SS}_{\mathrm{noise}}.
\end{equation}

In this additive ANOVA formulation, GCM and RCM effects are assumed to be independent. Any interaction between GCMs and RCMs, as well as other unexplained sources of variability, is therefore absorbed into the residual term. Any nonlinear dependencies between GCM and RCM, as well as any other departures from the additive structure, are captured implicitly in the residual term.

The relative contribution of each source of variability is quantified as the ratio of the corresponding sum of squares to the total sum of squares, \ie,
\begin{equation}
f_k = \frac{\mathrm{SS}_k}{\mathrm{SS}_{\mathrm{total}}},
\end{equation}
where $k \in \{\mathrm{GCM}, \mathrm{RCM}, \mathrm{res}\}$.

To compare the relative importance of GCM and RCM choices, we focus on the ratio of the GCM and RCM contributions to the total variability, expressed as
\begin{equation}
R = \frac{\mathrm{SS}_{\mathrm{RCM}}}{\mathrm{SS}_{\mathrm{GCM}}}.
\end{equation}

\subsubsection{Uncertainty partitioning} 
In this study, our ANOVA-based approach focuses specifically on \textit{model uncertainty}; that is, the variability arising from the choice of global and regional climate models. By decomposing the contributions of GCMs and RCMs to different climate aspects, we quantify how model choice drives the spread in projections. It is important to note that this approach does not explicitly model other sources of uncertainty. In particular, we do not explicitly account for \textit{scenario uncertainty}, which arises from unknown future greenhouse gas emissions. Furthermore, we do not treat \textit{internal climate variability}, which represents the natural, chaotic fluctuations of the climate system \citep{lehner2020partitioning}, as a separate component. 

\begin{figure}[h!]
    \centering
    \includegraphics[width=0.75\linewidth]{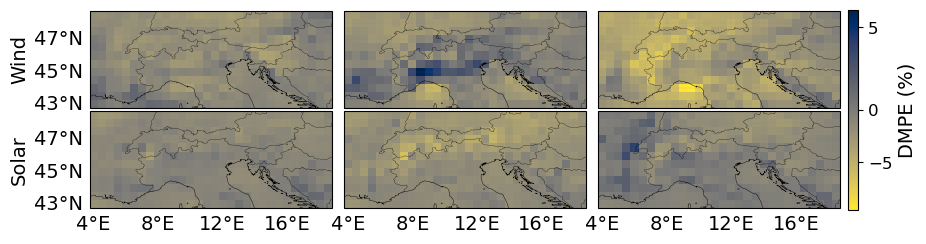}
    \caption{Spatial mean changes for the SMHI/MPI-M-MPI-ESM-LR ensemble between the 1995–2004 and 2045–2054 periods. Each map represents an individual ensemble member showing the difference in average fields. }
    \label{fig:internal-var}
\end{figure}

Instead, internal variability is implicitly distributed across the other components. Since each GCM–RCM combination represents a single realization, the measured signal for a specific model is inherently a composite of its physical response and internal noise. While utilizing 31 combinations dampens these stochastic effects, the variance fractions attributed to the GCM and RCM components still partially include internal variability. To mitigate this, we average the data over 10-year periods to filter out high-frequency internal noise. However, \Cref{fig:internal-var} shows that internal variability is substantial, even for 10- or 20 year (not shown) mean changes. 

Consequently, any uncertainty that cannot be explained by the main effects is captured within the residual sum of squares ($\mathrm{SS}_{\mathrm{res}}$). Crucially, because our matrix lacks multiple realizations for each GCM–RCM pair, $\mathrm{SS}_{\mathrm{res}}$	
includes RCM-GCM interactions and remaining stochastic noise. In contrast to the more complex sub-sampling methods proposed by \citet{hingray2014partitioning} to isolate internal variability, our approach only uses temporal filtering alongside the ANOVA decomposition. This provides a clear, interpretable framework for quantifying how the selection of GCMs and RCMs contributes to uncertainty in wind and solar projections.

\section{Results}
\subsection{Variability in climate data}
\label{sec:var_raw}
\begin{figure}[H]
    \centering
    \includegraphics[width=\linewidth]{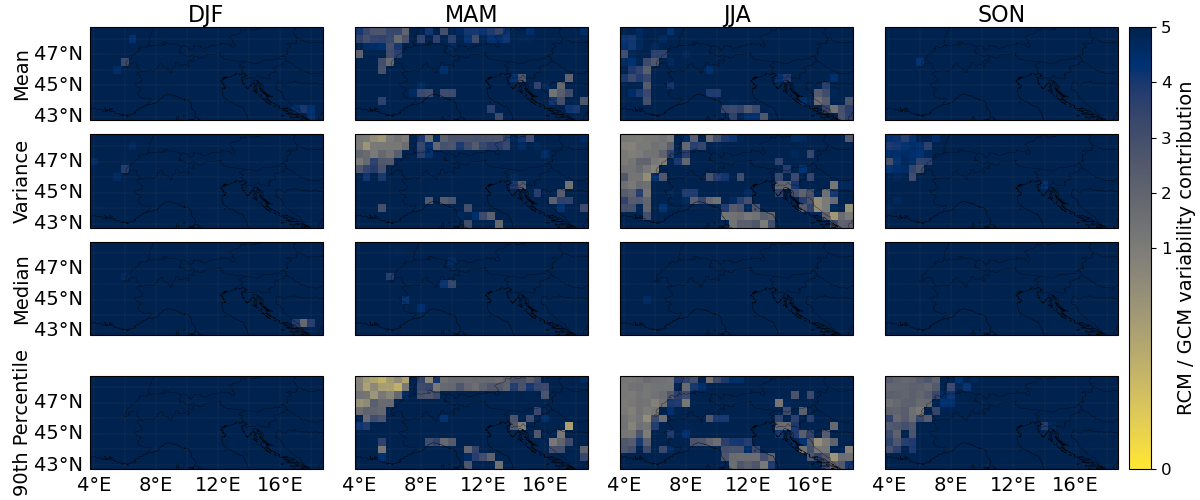}
    \caption{Results of the two-way ANOVA on historical solar radiation data between 1995-2004 across the 31 RCM–GCM pairs. The figure shows the ratio between RCM and GCM contribution to variability in different aspects of solar radiation, including seasonal mean, variance, median (50th percentile), and high extremes (90th percentile). RCMs dominate variability in most regions, seasonal differences are also visible.}
    \label{fig:overview-solar-raw}
\end{figure}

\begin{figure}[H]
    \centering
    \includegraphics[width=\linewidth]{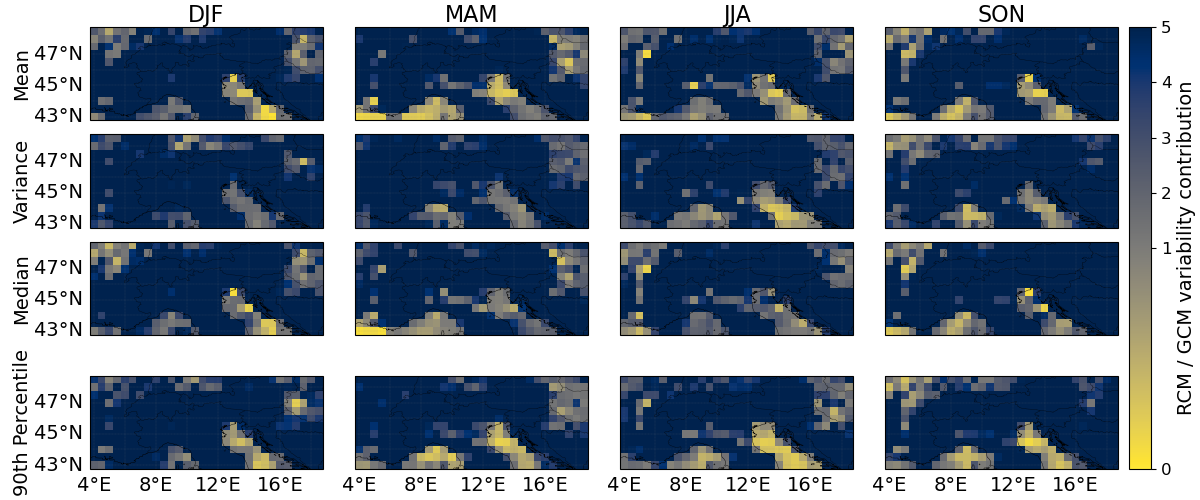}
    \caption{Results of the two-way ANOVA on historical wind speed data between 1995-2004 across the 31 RCM–GCM pairs. The figure shows the ratio between RCM and GCM contribution to variability in different aspects of surface wind speed, including seasonal mean, variance, median (50th percentile), and high extremes (90th percentile). RCMs dominate variability in most regions, GCMs dominate over seas.}
    \label{fig:overview-wind-raw}
\end{figure}

We investigate the sources of variability by performing ANOVA across RCMs and GCMs. For raw solar radiation data, we find that most variability in the study region across single-variable aspects originates from RCMs; the results are shown in \Cref{fig:overview-solar-raw}. Additionally, the data show a slight seasonal trend in the mean, variance, and 90th percentile; this trend is not persistent in the median. 
In contrast, \Cref{fig:overview-wind-raw} shows that wind speed variability exhibits strong regional differences: RCMs contribute more over land, particularly near the Alps, while GCMs dominate over the sea. The uncertainty partitioning remains highly consistent across the different aspects, with no substantial differences emerging between the mean, variance, or percentiles.

\begin{figure}[H]
    \centering
    \includegraphics[width=\linewidth]{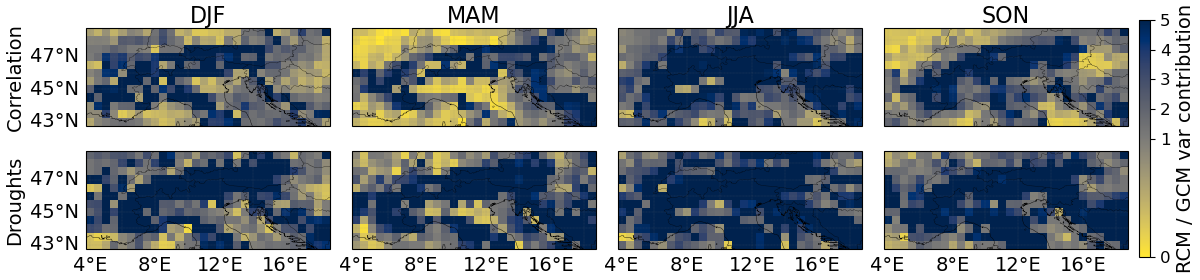}
    \caption{Results of the two-way ANOVA on combined aspects of solar radiation and wind speed data between 1995-2004 across the 31 RCM–GCM pairs. The figure shows the ratio between RCM and GCM contribution to variability in correlation and drought days. RCMs dominate variability in most regions, particularly in high elevation regions.}
    \label{fig:overview-combined-raw}
\end{figure}

The results for the combined aspects (\Cref{fig:overview-combined-raw}) are similar to those for wind. In particular, in mountainous Alpine regions, the variability in spatial correlation and drought events is dominated by RCM effects. However, the combined aspects exhibit a seasonal pattern, with greater variability attributable to RCMs in winter than in summer. Notably, the relative contribution of RCMs to the total variance is higher for these combined aspects than for single-variable aspects. 


\subsection{Variability of climate change}
\label{sec:var-change}
\begin{figure}[H]
    \centering
    \includegraphics[width=\linewidth]{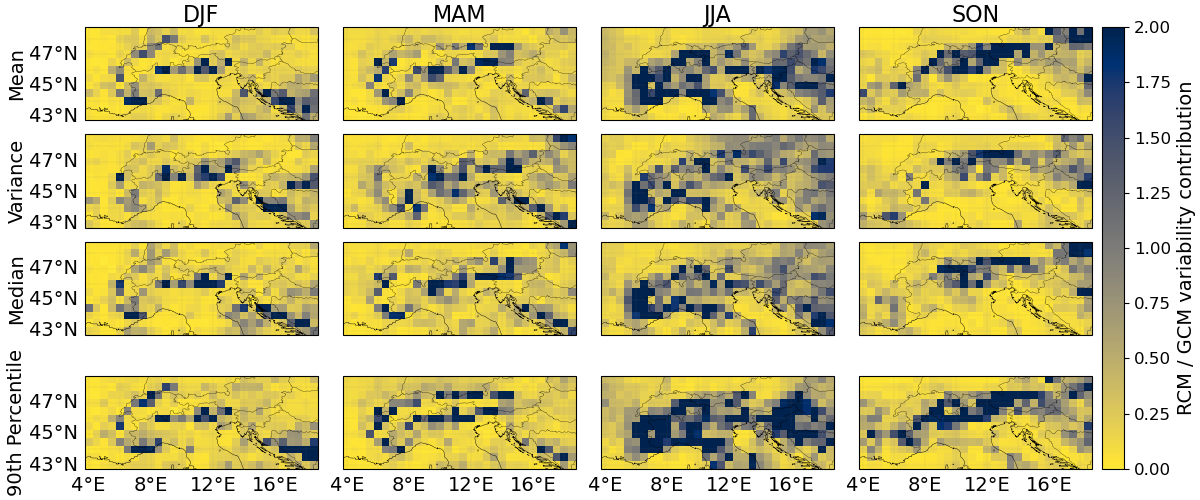}
    \caption{Results of the two-way ANOVA on changes of wind speed data between 1995-2004 and 2045-2054. The figure shows the ratio between RCM and GCM contribution to variability in different aspects of solar radiation changes, including seasonal mean, variance, median (50th percentile), and high extremes (90th percentile). RCMs dominate variability close to the Alps, whereas GCMs dominate in flat regions.}
    \label{fig:overview-wind-change}
\end{figure}

We now apply the same analysis to the projected changes in the different aspects; see \Cref{fig:overview-wind-change}. For wind, the results are structurally similar to those from the raw data. However, we find that changes are much more dominated by GCMs, and RCMs only contribute more to the variability in the narrow Alpine region. Compared to the raw data, the influence of RCMs is weaker, and the area where they dominate is considerably smaller. The differences between individual aspects are small. 

\begin{figure}[H]
    \centering
    \includegraphics[width=\linewidth]{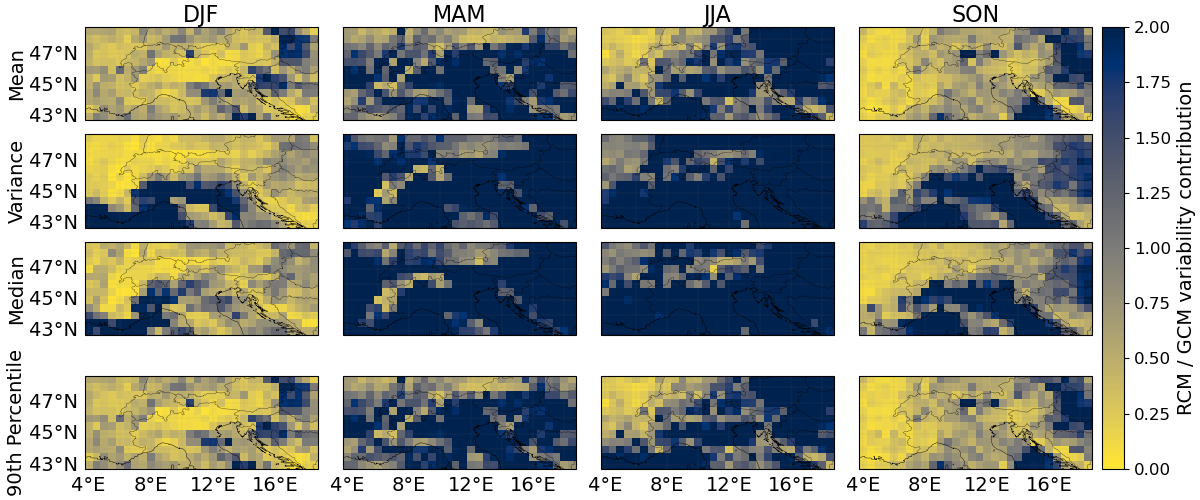}
    \caption{Results of the two-way ANOVA on changes of solar radiation data between 1995-2004 and 2045-2054. The figure shows the ratio between RCM and GCM contribution to variability in different aspects of solar radiation changes, including seasonal mean, variance, median (50th percentile), and high extremes (90th percentile). RCMs dominate variability primarily in summer, whereas GCMs dominate in winter.}
    \label{fig:overview-solar-change}
\end{figure}

For changes in solar radiation (\Cref{fig:overview-solar-change}), the pattern differs from the raw solar radiation data. While variability in the raw data is largely dominated by RCMs, the variability in projected changes exhibits pronounced seasonal differences. RCMs dominate most of the variability in spring and summer, whereas GCMs contribute more strongly in autumn and winter. While differences between the individual aspects are generally negligible, a distinct divergence emerges in the spatial structures of the 90th percentile.

\begin{figure}[H]
    \centering
    \includegraphics[width=\linewidth]{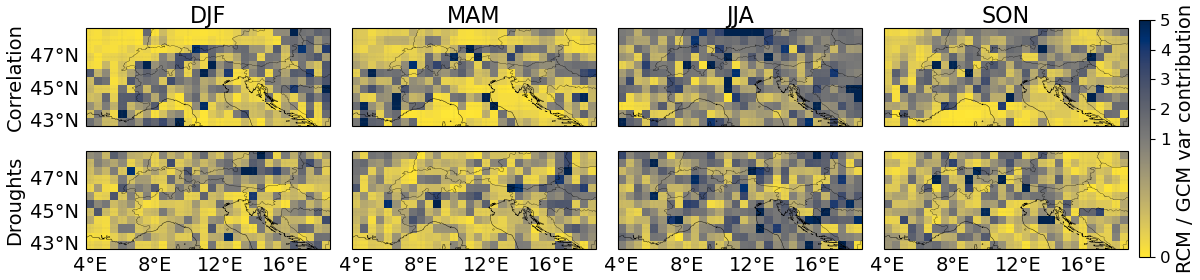}
    \caption{Results of the two-way ANOVA on changes of combined aspects between 1995-2004 and 2045-2054. The figure shows the ratio between RCM and GCM contribution to variability in correlation and drought days. GCMs dominate variability in most regions, with slightly more RCM contribution in summer.}
    \label{fig:overview-combined-change}
\end{figure}

In the combined case (\Cref{fig:overview-combined-change}), spatial patterns are less pronounced. For spatial correlation, variance in the Alpine region is dominated by RCMs, whereas most other regions are dominated by GCMs, particularly in winter and spring. In summer, no clear spatial patterns emerge. In all other seasons, less than one third of the area is dominated by RCMs. During drought days, spatial patterns are even less pronounced. Across all seasons, variability is dominated by GCMs. Nevertheless, the general tendency for stronger RCM influence in summer is also evident here, with 55\% of the region being more strongly influenced by GCMs in summer compared to 78\% in winter.

\subsubsection{Robustness with respect to data selection}
We repeated the same analyses using two different selection methods and present the results in \Cref{sec:subset}. The results remain consistent across both selection methods, with only minor variations in the local values, suggesting that our findings are not dependent on a specific data subset.

\subsection{Projected changes}
\begin{figure}[H]
    \centering
    \includegraphics[width=0.75\linewidth]{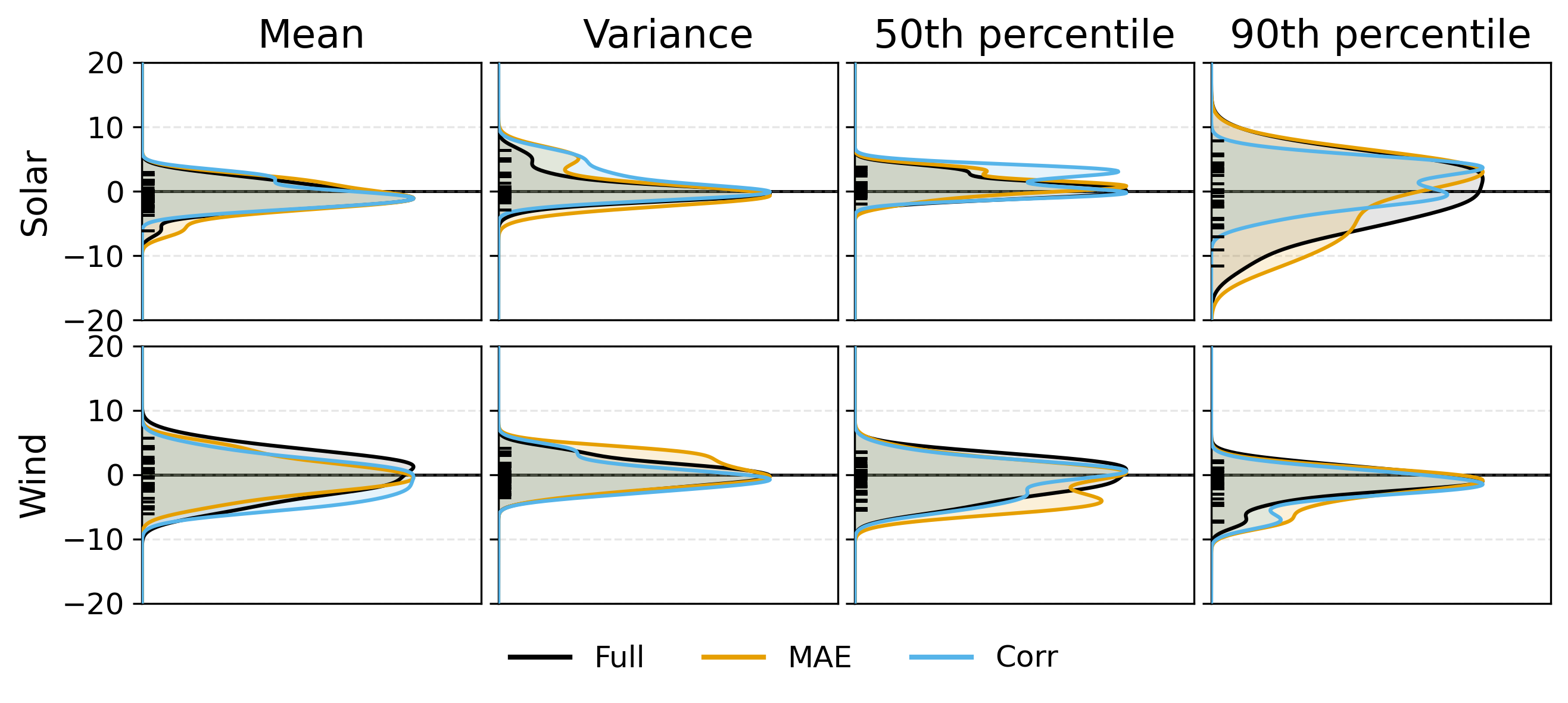}
    \caption{Average shifts in solar and wind energy metrics. Kernel density estimations (KDEs) of the ensemble distribution of regional average changes between the historical (1995–2004) and future (2045–2054) periods. The vertical axis denotes the percentage change, with a horizontal reference line at zero. Black lines along the y-axis indicate the shift projected by individual ensemble members, members above the line project an increase in the spatial mean. Colored distributions highlight subsets selected by mean absolute error (MAE, orange) and temporal correlation (Corr, blue) relative to the full ensemble (black). }
    \label{fig:0068_change_pdf_combined}
\end{figure}

The variability identified in the ANOVAs translates to a spread of regional projections. Notably, the presence of a spread in distributions does not automatically imply a lack of agreement on the sign of the change. However, our results indicate that the direction of the shifts here lacks consistency. The distributions in \Cref{fig:0068_change_pdf_combined} represent the spread of spatially averaged climate change signals; \ie the difference between future and historical time periods. We find that, for most aspects, the projected changes are split almost evenly between positive and negative directions. Furthermore, data selection typically does not reduce the overall spread of these projected average changes, with the direction of change remaining similarly varied.

This regional-scale spread is mirrored at the grid-point level. We assess the robustness of projected changes and find that only a few locations exhibit consistent trends across models. Following the IPCC definition, a shift is considered robust if it occurs in at least 80\% of models and exceeds internal variability in at least 66\% of cases \citep{lee2021future}. Here, we focus on the first criterion and observe that for most locations, predicted changes are inconsistent. Detailed results and location-specific assessments are provided in Appendix \Cref{sec:location-specific}.

Our results indicate that a substantial part of the region may be affected by climate change (\Cref{fig:0068_change_pdf_combined}), with large associated uncertainties. However, a comparison between the location-specific changes (\Cref{sec:location-specific}) and the identified variability patterns (\Cref{sec:var_raw} and \Cref{sec:var-change}) reveals no identifiable similarities.

\section{Discussion}
\label{sec:discussion}
This study provides a spatially and seasonally resolved decomposition of the uncertainty in raw data and projected changes in wind and solar resources, disentangling the variability contribution from GCMs and RCMs. The results highlight that uncertainty patterns are not uniform across variables or seasons. For wind, large-scale circulation patterns dominate the response in many regions, leading to a strong impact of the driving GCMs across all seasons. In contrast, solar radiation changes show a seasonal cycle and are, particularly in the summer months, more sensitive to RCM formulation. These variable-specific sensitivities produce distinct spatial patterns in the relative contributions of GCMs and RCMs that cannot be captured by global models alone. Furthermore, the uncertainty partitioning remains largely consistent, regardless of whether changes are evaluated for the mean, median, variance, or 90th percentile. Notably, the patterns of change and variability do not align, suggesting that the underlying source of uncertainty may not be a reliable indicator of whether models will agree on the direction of change.

Our results indicate that GCMs are generally sufficient for assessing the uncertainty of the impacts of climate change on wind power over flat regions, consistent with \eg \citet{rummukainen2016added}, who note that over smooth terrain, RCMs have limited influence on the results. However, our findings again highlight the importance of RCMs in mountainous regions. In addition, we show that RCMs contribute substantially to uncertainty in solar radiation in spring and summer, underscoring the need to exploit RCMs in solar power studies, particularly given the higher share of solar power generation during the summer months in Central Europe \citep[\eg][]{heide2010seasonal}. In the combined analyses, variability is dominated by GCMs, especially in winter, when electricity demand is highest \citep{wenz2017north} and wind power plays a central role in supply.

Overall, given potentially shifting electricity demand patterns \citep{wenz2017north}, a diverse set of GCMs and RCMs is required to ensure robust assessments of renewable energy resources. This underscores the necessity of high-resolution regional climate simulations or statistical alternatives, such as emulators \citep{effenberger2025bridging}. Because climate impacts vary by energy system design \citep{bloin2025climate}, future assessments should simultaneously address climate and energy system uncertainties.

Previous studies found that uncertainties in climate projections have important implications for renewable energy applications \citep{yang2022climate}. In our study, the dominance of different sources of uncertainty varies depending on the region and season, complicating the use of climate projections in energy system planning. For wind, reducing uncertainty will likely require model improvements in global circulation patterns, and quantifying uncertainty requires exploring a large GCM-RCM matrix. For solar radiation, progress in cloud representation and surface processes at regional scales may be more critical--particularly in summer, when solar power plays a more important role than in winter. For solar radiation studies, it will therefore be of greater importance to exploit many different RCMs, as more GCMs add less information regarding uncertainty quantification. Ensemble-based approaches, including both multiple RCMs and GCMs, remain essential for risk assessment and long-term planning in power systems.

In this study, we quantify model uncertainty in wind and solar projections by analyzing the variability across an ensemble of GCM–RCM pairs. We interpret this variance as a proxy for uncertainty, capturing differences in model outputs due to differences in GCMs and RCMs. ANOVA allows us to decompose the contributions of each model type to the variability in historical fields and projected changes. However, this approach does not account for other important sources of uncertainty in climate data, such as internal climate variability \citep{wohland2021wind} or scenario uncertainty related to future emissions \citep{russo2023future}, or uncertainties arising from energy system configurations \citep{bloin2025climate} and demand changes \citep{wenz2017north}. As such, the variability quantified here provides a lower-bound estimate of total uncertainty, reflecting only the component attributable to differences between the climate models themselves.

ANOVA provides a clear and interpretable decomposition. However, we do not have access to the full RCM-GCM matrix and, therefore, choose not to model the combined effects. An alternative approach would be to impute the missing data, which has been adopted in other studies, \eg \citep{christensen2022filling, deque2012spread}. One shortcoming of ANOVA is that the analysis implicitly assumes the linear separability of GCM and RCM effects, whereas climate processes may involve higher-order interactions \citep{christensen2022filling, evin2021balanced}. Furthermore, the internal variability of the individual models is not explicitly addressed, although it can be substantial at regional scales (compare \Cref{fig:internal-var}), particularly for wind \citep{wohland2021wind}. Final results are inherently dependent on the selected ensemble \citep{lira2025assessing}. Nevertheless, our sensitivity analysis in \cref{sec:subset} confirms that the results are robust to different data selection methods, suggesting that the observed contributions are consistent across model subsets.

Our analysis provides a systematic framework for identifying the drivers of uncertainty in wind and solar projections within multi-model ensembles. These findings support a more informed application of climate data in renewable energy planning and highlight priority areas for model refinement. While this study focuses on the climatically diverse Alpine region, future research should expand this methodology to other geographical contexts. Furthermore, investigating how these uncertainties translate into impacts remains a critical next step.

\section{Conclusion}
We provide a spatially and seasonally resolved assessment of uncertainty in wind and solar climate projections by decomposing contributions from GCMs and RCMs. When considering changes over time, wind variability is largely driven by GCMs, with RCMs adding value mainly in mountainous regions, while solar radiation shows a strong seasonal dependence, with RCMs dominating in spring and summer. Combined analyses highlight that GCMs dominate winter variability, when electricity demand peaks, emphasizing the importance of considering both climate and demand patterns. Our results demonstrate that a robust uncertainty assessment requires diverse ensembles of GCMs and RCMs, particularly given the additional uncertainties associated with demand and energy system configurations highlighted in related work. These findings provide guidance for designing climate model ensembles for renewable energy applications and highlight areas for model improvement.

\subsection*{Acknowledgments}
We thank Luca Schmidt for her helpful feedback on earlier drafts of the manuscript. 
\bibliographystyle{plainnat}
\bibliography{sample}
\newpage
\newpage
\begin{appendices}
\section{Subset experiments}
\label{sec:subset}
\definecolor{wind_corr}{RGB}{0,0,0}    
\definecolor{wind_mape}{RGB}{230,159,0}      
\definecolor{solar_mape}{RGB}{86, 180, 233}   
\definecolor{solar_corr}{RGB}{0,158,115}   

\begin{table}[h]
    \centering
    \small
    \caption{EURO-CORDEX RCM–GCM simulation pairs were selected based on correlation and mean absolute percentage error relative to ERA5. \textcolor{solar_corr}{\checkmark} marks models with high correlation for solar radiation, \textcolor{wind_corr}{\checkmark} those with high correlation for wind speed. \textcolor{solar_mape}{\checkmark} indicates low MAPE for solar radiation, and \textcolor{wind_mape}{\checkmark} low MAPE for wind speed. All datasets, except one, appear in at least one of the subsets.}
    \label{tab:dataselection}
\begin{tabularx}{\textwidth}{l|*{6}{>{\centering\arraybackslash}X}}
\hline
RCM $\backslash$ GCM 
& CNRM-CERFACS-CNRM-CM5 
& ICHEC-EC-EARTH 
& IPSL-IPSL-CM5A-MR 
& MOHC-HadGEM2-ES 
& MPI-M-MPI-ESM-LR 
& NCC-NorESM1-M \\
\hline
CLMcom-ETH &\textcolor{solar_mape}{\checkmark}&\textcolor{solar_mape}{\checkmark}  &$\times$&\textcolor{solar_mape}{\checkmark}  & \textcolor{wind_mape}{\checkmark}\textcolor{solar_mape}{\checkmark} & \textcolor{wind_corr}{\checkmark}\textcolor{solar_mape}{\checkmark} \\
CNRM       &\textcolor{solar_corr}{\checkmark} &$\times$&$\times$& \textcolor{solar_corr}{\checkmark}&\textcolor{solar_corr}{\checkmark}& \textcolor{wind_corr}{\checkmark}\textcolor{solar_corr}{\checkmark} \\
DMI        & \textcolor{wind_corr}{\checkmark}\textcolor{wind_mape}{\checkmark}\textcolor{solar_mape}{\checkmark} & \textcolor{wind_corr}{\checkmark}\textcolor{wind_mape}{\checkmark} & \textcolor{wind_corr}{\checkmark} \textcolor{wind_mape}{\checkmark}& \textcolor{wind_corr}{\checkmark}\textcolor{wind_mape}{\checkmark}\textcolor{solar_corr}{\checkmark}& \textcolor{wind_corr}{\checkmark}\textcolor{wind_mape}{\checkmark} & \textcolor{wind_corr}{\checkmark}\textcolor{wind_mape}{\checkmark} \\
KNMI       &\textcolor{solar_mape}{\checkmark}& \textcolor{solar_mape}{\checkmark} &\textcolor{wind_mape}{\checkmark}\textcolor{solar_mape}{\checkmark} &  & \textcolor{wind_mape}{\checkmark}\textcolor{solar_mape}{\checkmark} & \textcolor{wind_corr}{\checkmark}\textcolor{wind_mape}{\checkmark}\textcolor{solar_mape}{\checkmark} \\
MOHC       &\textcolor{solar_mape}{\checkmark}\textcolor{solar_corr}{\checkmark} &$\times$&$\times$&\textcolor{solar_corr}{\checkmark} &  \textcolor{solar_mape}{\checkmark}\textcolor{solar_corr}{\checkmark}& \textcolor{wind_corr}{\checkmark}\textcolor{solar_mape}{\checkmark}\textcolor{solar_corr}{\checkmark}\\
SMHI       & \textcolor{wind_corr}{\checkmark}\textcolor{wind_mape}{\checkmark}\textcolor{solar_corr}{\checkmark} & \textcolor{wind_corr}{\checkmark} \textcolor{wind_mape}{\checkmark}\textcolor{solar_corr}{\checkmark}& \textcolor{wind_corr}{\checkmark}\textcolor{wind_mape}{\checkmark}\textcolor{solar_corr}{\checkmark} & \textcolor{wind_corr}{\checkmark}\textcolor{wind_mape}{\checkmark}\textcolor{solar_corr}{\checkmark} & \textcolor{wind_corr}{\checkmark}\textcolor{wind_mape}{\checkmark}\textcolor{solar_corr}{\checkmark}&\textcolor{solar_corr}{\checkmark}  \\
\hline
\end{tabularx}
\end{table}
\counterwithin{figure}{section}
\counterwithin{table}{section}
\begin{figure}[H]
    \centering
    \includegraphics[width=\linewidth]{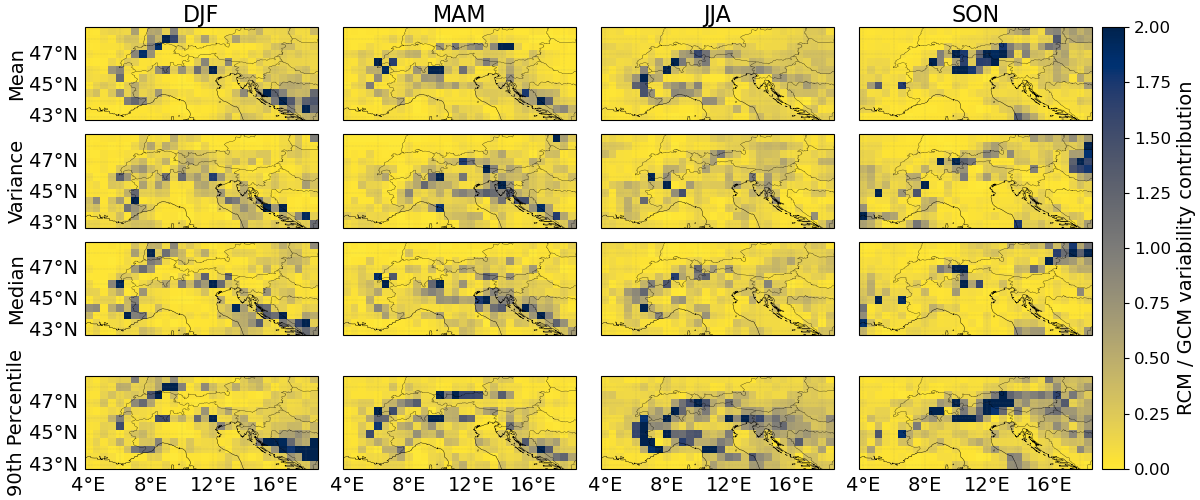}
    \caption{Results of the two-way ANOVA on changes of wind speed data between 1995-2004 and 2045-2054. The data used corresponds to the 15 datasets with lowest MAPE. The figure shows the ratio between RCM and GCM contribution to variability in different aspects of solar radiation changes, including seasonal mean, variance, median (50th percentile), and high extremes (90th percentile). RCMs dominate variability close to the Alps, whereas GCMs dominate in flat regions.}
    \label{app:overview-wind-change}
\end{figure}
\begin{figure}[H]
    \centering
    \includegraphics[width=\linewidth]{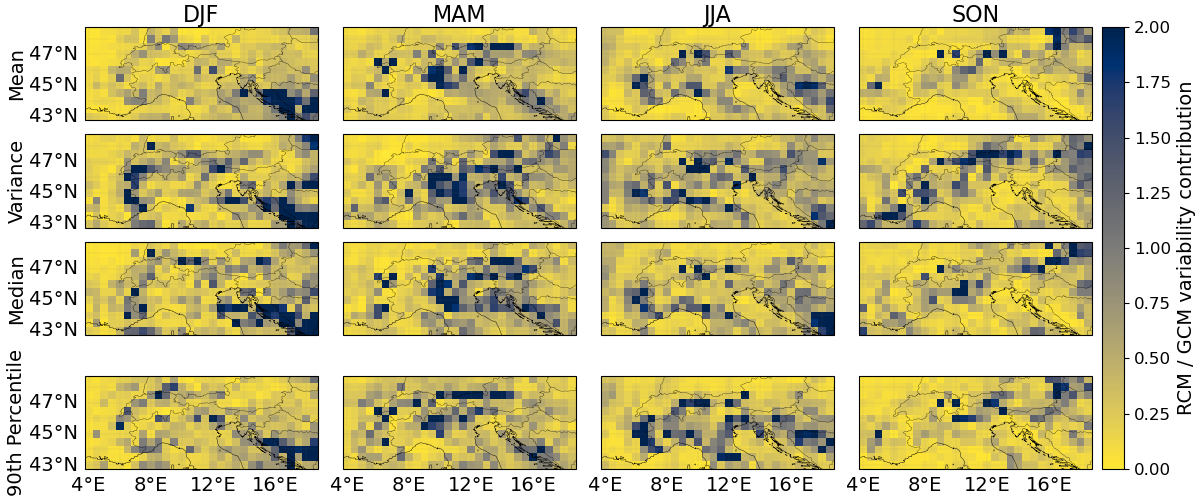}
    \caption{Results of the two-way ANOVA on changes of wind speed data between 1995-2004 and 2045-2054. The data used corresponds to the 15 datasets with highest correlation. The figure shows the ratio between RCM and GCM contribution to variability in different aspects of solar radiation changes, including seasonal mean, variance, median (50th percentile), and high extremes (90th percentile). RCMs dominate variability close to the Alps, whereas GCMs dominate in flat regions.}
    \label{app:overview-wind-change}
\end{figure}
\begin{figure}[H]
    \centering
        \includegraphics[width=\linewidth]{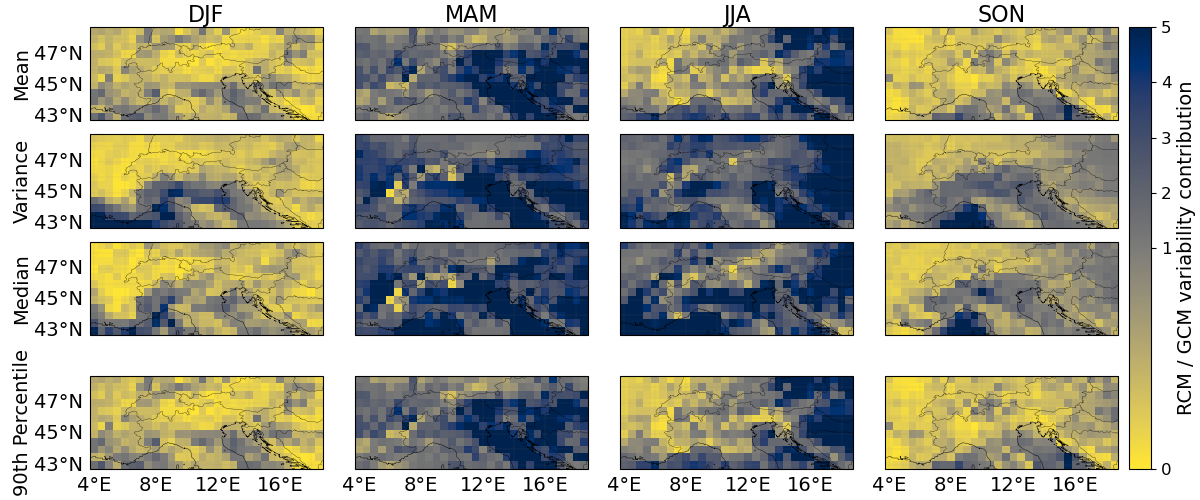}
    \caption{Results of the two-way ANOVA on changes of solar radiation data between 1995-2004 and 2045-2054. The data used corresponds to the 15 datasets with lowest MAPE. The figure shows the ratio between RCM and GCM contribution to variability in different aspects of solar radiation changes, including seasonal mean, variance, median (50th percentile), and high extremes (90th percentile). RCMs dominate variability close to the Alps, whereas GCMs dominate in flat regions.}
    \label{app:overview-solar-change}
\end{figure}
\begin{figure}[H]
    \centering
    \includegraphics[width=\linewidth]{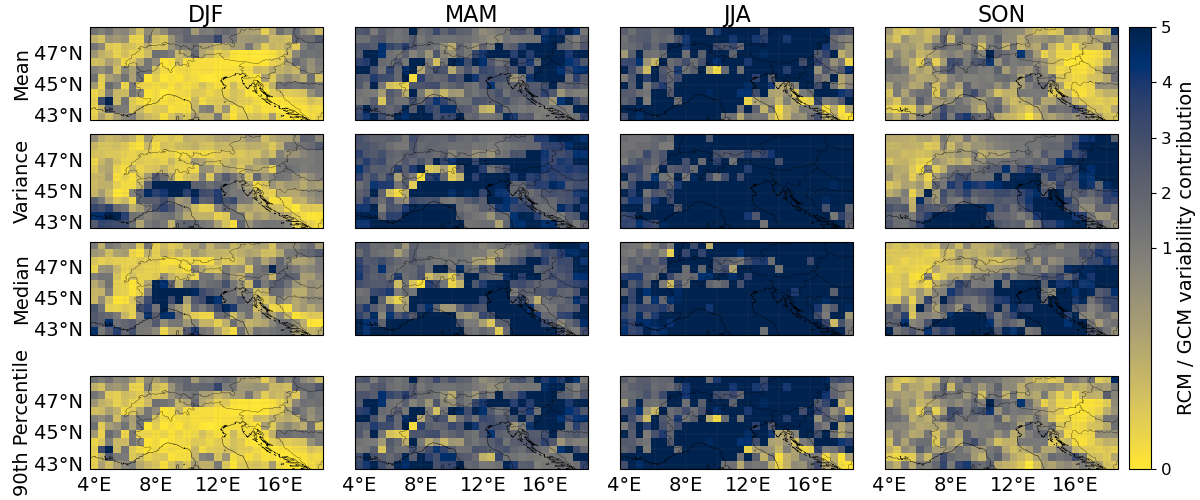}
    \caption{Results of the two-way ANOVA on changes of solar radiation data between 1995-2004 and 2045-2054. The data used corresponds to the 15 datasets with highest correlation. The figure shows the ratio between RCM and GCM contribution to variability in different aspects of solar radiation changes, including seasonal mean, variance, median (50th percentile), and high extremes (90th percentile). RCMs dominate variability close to the Alps, whereas GCMs dominate in flat regions.}
    \label{app:overview-solar-change}
\end{figure}
\newpage
\section{Location-specific changes}
\label{sec:location-specific}
\begin{figure}[H]
    \centering
    \includegraphics[width=0.9\linewidth]{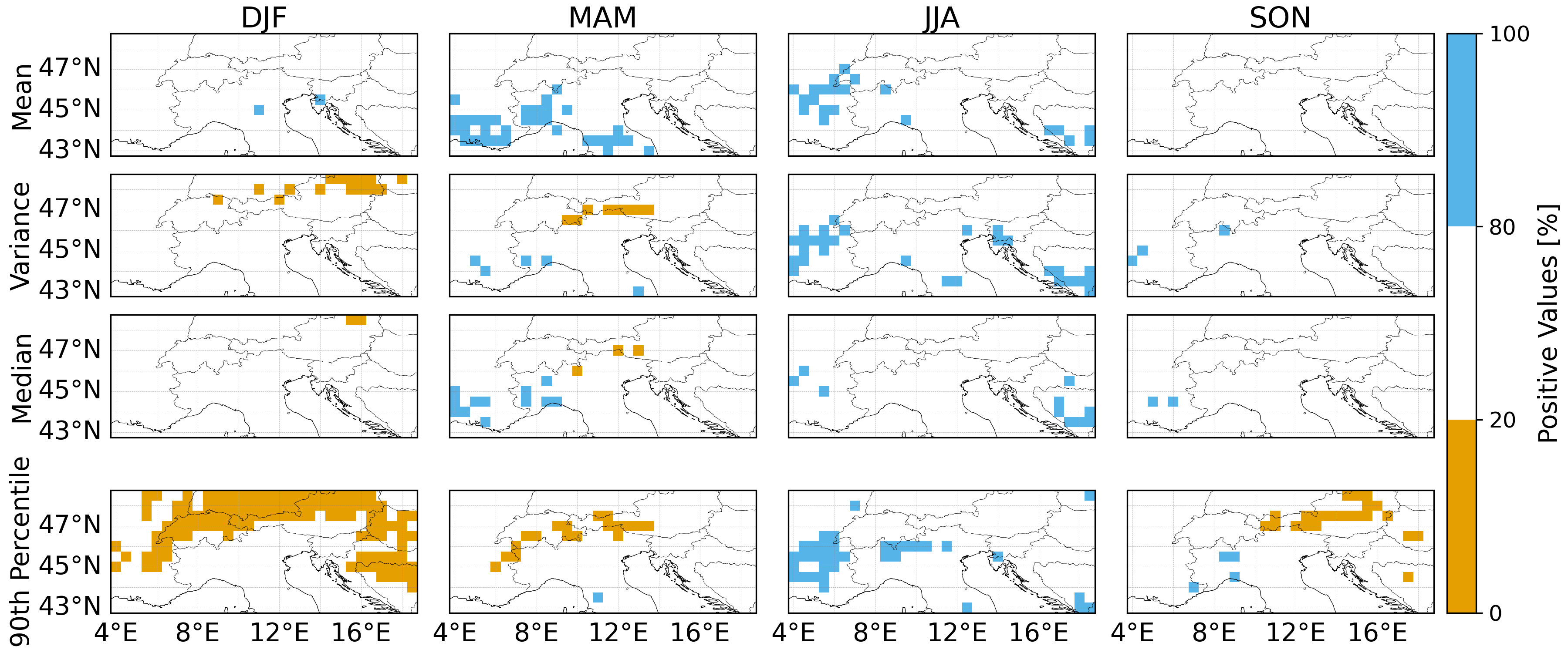}
    \caption{Projected changes in solar radiation for each location, showing the four analyzed aspects (mean, variance, 50th percentile, and 90th percentile). Blue indicates an increase in the future period (2045–2054) relative to the historical period (1995–2004), while orange indicates a decrease. }
    \label{fig:overview-wind}
\end{figure}
\begin{figure}[H]
    \centering
    \includegraphics[width=0.9\linewidth]{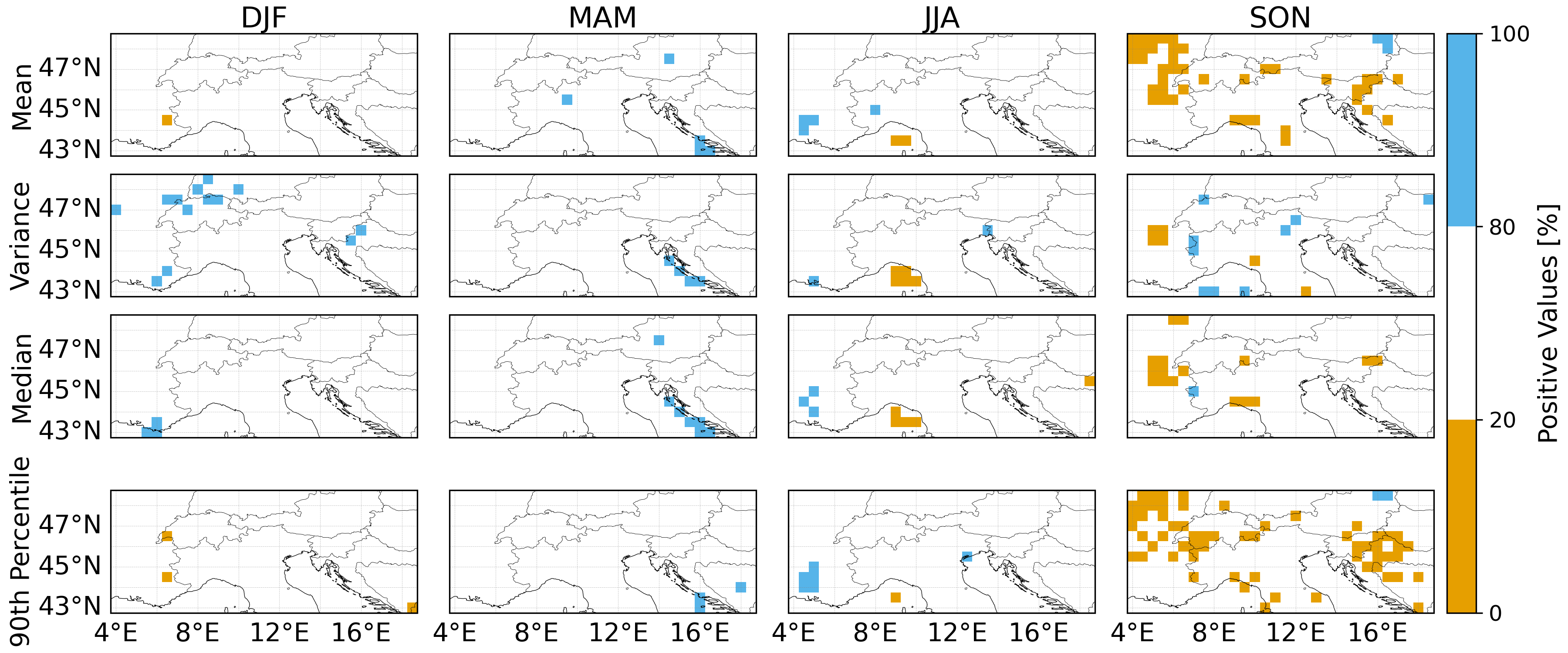}
    \caption{Projected changes in wind speed for each location, showing the four analyzed aspects (mean, variance, 50th percentile, and 90th percentile). Blue indicates an increase in the future period (2045–2054) relative to the historical period (1995–2004), while orange indicates a decrease.}
    \label{fig:overview-wind}
\end{figure}
\begin{figure}[H]
    \centering
    \includegraphics[width=0.9\linewidth]{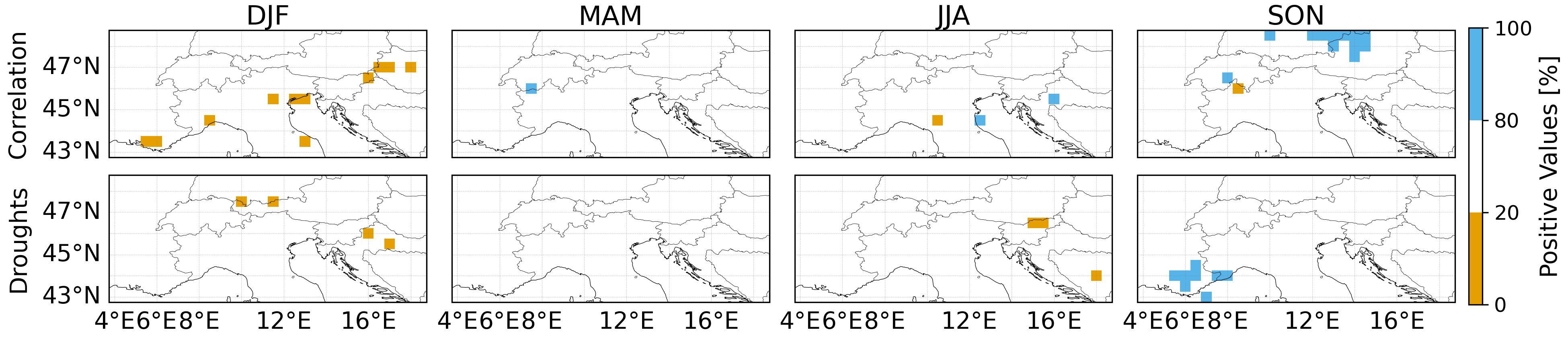}
    \caption{Projected changes in combined aspects for each location (correlation, drought days). Blue indicates an increase in the future period (2045–2054) relative to the historical period (1995–2004), while orange indicates a decrease.}
    \label{fig:overview-solar}
\end{figure}
\end{appendices}
\end{document}